\documentstyle[12pt,a4wide]{article}

\def\a{\alpha}  \def\s{\sigma} \def\da{{\dot{\a}}}
\def\ds{{\dot{\s}}}  \def\b{\beta}

\newcommand{\plabel}[1]{\label{#1}}
\newcommand{\p}[1]{(\ref{#1})}
\newcommand{\bi}[1]{\bibitem{#1}}

\begin{document}

\title{Matrix Supermultiplet of $N=2$, $D=4$ Supersymmetry
 and Supersymmetric 3-brane}
\author{ A. Kapustnikov\thanks{\it E-mail: kpstnkv@creator.dp.ua}
\ and A. Shcherbakov\thanks{\it E-mail: novel@ff.dsu.dp.ua} \\
\it Department of Physics, Dnepropetrovsk University, \\
49050, Dnepropetrovsk, Ukraine}

\date{}
\maketitle
\begin{abstract}
It is shown that the Lagrangian density of the supersymmetric 3-brane
can be regarded as a component of an infinite-dimensional
supermultiplet of $N=2$, $D=4$ supersymmetry spontaneously broken
down to $N=1$. The latter is described by $N=1$ Hermitian bosonic
matrix superfield $V_{mn} = V^\dagger_{nm}$, $[V_{mn}] = m+n$, $m,n
= 0,1,\ldots$ in which the component $V_{01}$ is identified with a
chiral Goldstone $N=1$ multiplet associated with central charge of
the $N=2$, $D=4$ superalgebra, and $V_{11}$ obeys a specific
nonlinear recursive equation providing the
possibility to express $V_{11}$ (as well as the other components
$V_{mn}$) covariantly in terms of $V_{01}$. We demonstrate that the
solution of $V_{11}$ gives the right \emph{PBGS} action for the
super-3-brane.
\end{abstract}

\section{Introduction}
Supersymmetric 3-brane gives a remarkable pattern of a theory where
the partial breaking of global supersymmetry \emph{(PBGS)} in four
spacetime dimensions is occurred \cite{hp,hlp}. The bosonic part of
the corresponding action amounts to the ``static gauge'' form of the
Nambu--Goto action which can be derived, for example, from the
nonlinear realization of $D=6$ relativistic symmetry in the coset space
$ISO(1,5)/SO(1,3)\times SO(2)$ \cite{w,i}. The supersymmetric
generalization of this approach is elaborated in ref. \cite{bg1}. There
the coset space $ISO(1,5)/SO(1,3)\times SO(2)$ is supposed to be
embedded into the supersymmetric one $G/H$ involving the superbrane
worldvolume coordinates $z = \{x, \theta, \bar{\theta}\}$ along with
the set of the Goldstone superfields $\{\phi(z)$, $\bar\phi(z)$,
$\psi_\a (z)$, $\bar{\psi}_{\da}(z)$, $\Lambda_a(z)$, $\Xi(z)\}$
associated with the generators of spontaneously broken symmetries
$\{Z=P_4 - iP_5$, $S_\a$, $\bar{S}_{\da}$, $K_a$, $T\}$. The main
advantage of this approach is that it allows one to find the action
in terms of the Goldstone superfields on the base of the standard
method of the nonlinear realization of supersymmetry in the coset
superspace $G/H$ \cite{cwz,v,o,ivk}. In the case of the $N=2
\rightarrow N=1$, $D=4$ \emph{PBGS} pattern it gives the superfield
action of the supersymmetric 3-brane propagating in the
six-dimensional Minkowski spacetime. The component form of this
action exactly coincides with the Green--Schwarz covariant action for
the supermembrane obtained in ref. \cite{hlp}. Note, that the final
expression of the superfield action depends only on the scalar
Goldstone superfields $\phi$, $\bar{\phi}$ and their covariant
derivatives. The superfields $\psi_\a (z)$, $\bar{\psi}_{\da} (z)$,
$\Lambda_a(z)$, $\Xi(z)$ are not involved due to the
covariant constraints imposed on the corresponding Cartan forms.
Nevertheless, this action is not very convenient for applications
since the superfields $\phi$, $\bar{\phi}$ themselves have to be
constrained by the nonlinear conditions $\bar{\nabla}_{\da}\phi =
\nabla_\a \bar{\phi} = 0$, where $\nabla_\a$, $\bar{\nabla}_{\da}$
are the spinor covariant derivatives of the nonlinear realization.
This makes the corresponding variation procedure rather complicated
since these objects themselves depend nonlinearly on $\phi$,
$\bar{\phi}$. The situation gets worse even more because so far we do
not know exactly a full non-perturbative \emph{PBGS} action of the
super-3-brane derived from the nonlinear realization. Today it is
specified only up to the fourth order in the superfields $\phi$,
$\bar{\phi}$ \cite{bg1}.

Fortunately, there exists another approach which significantly
simplifies these computation problems. In contrast to the previous
approach to the system under consideration this one is based on the
linear realization of supersymmetry \cite{bg2,rt}. It strongly
resembles the \emph{PBGS} theory of the supersymmetric $D3$-brane
(see, e.g. \cite{bg3}), where the hidden supersymmetry is described
by the $N=2$, $D=4$ vector supermultiplet. It turns out that in the
case of the supersymmetric 3-brane the corresponding Goldstone
supermultiplet can be described by the $N=2$ tensor supermultiplet
constrained by the suitable nilpotence conditions \cite{rt}. Solving
them we get the south for action in terms
of the \emph{linear} $N=1$ Goldstone superfield. It was shown in ref. \cite{rt}
that this action can be equivalently transformed into the chiral
Goldstone superfield action \cite{bg2} with the help of a
chiral--linear superfield duality transformation. This fact, indeed,
radically simplifies the computation problems since in this case we
should vary only the unconstrained superfield of the linear
realization $\phi(x + 2i\theta\bar{\theta},~\theta)$. At last, the
obvious advantage of this approach as compared with the
aforementioned one is that it provides the possibility to get a full
non-perturbative form action of the super-3-brane.

Nevertheless, there is a problem which remains unsolved even in the
framework of this approach. Actually, from the papers \cite{bg2,rt} we
know how to construct the Lagrangian density of the super-3-brane in
terms of the \emph{linear} $N=1$ Goldstone superfield and how to
transform it with the help of the duality transformations into the
final form action depending on the \emph{chiral} Goldstone
superfield $\phi$.
However, implementing such a procedure we inevitably lose a
straightforward contact between the Lagrangian and its transformation
law with respect to the $N=2$, $D=4$ supersymmetry. Precisely, we do
not know today how explicitly it transforms with respect to the
variation of $\phi$?

In this article we are going to show that there exists a simple
answer to this question. In fact, the required transformation law
can be derived from the new infinite-dimensional matrix $N=2$ supermultiplet
which incorporates both the chiral Goldstone multiplet $\phi$
and the Lagrangian density ${\cal L}$ as its independent
scalar $N=1$ components.

\section{Extended vector $N=2$, $D=4$ supermultiplet}
Let us begin with the central charge extension of the $N=2$, $D=4$
Poincar{\'e} superalgebra
\begin{eqnarray} \plabel{alg}
\{Q_{\a},\bar{Q}_{\dot{\a}}\} &=& -2P_{\a\da},\qquad
\{S_{\a},\bar{S}_{\dot{\a}}\} = -2P_{\a\da}, \\
\{Q_{\a},S_{\b}\} &=& 2\varepsilon_{\a\b}Z,~\qquad
\{\bar{Q}_{\dot{\a}},\bar{S}_{\dot{\b}}\} =
2\varepsilon_{\dot{\a}\dot{\b}}\bar{Z},
\nonumber
\end{eqnarray}
where the other (anti)commutators are supposed to vanish. In what follows
the $N=2$ superspace notations for the generators will be used
\begin{eqnarray} \plabel{gens}
&&P_{\a\da}=i\partial_{\a\da},\quad
    Q_{\a}=\partial^{(\theta)}_{\a} +
        i\bar{\theta}^{\dot{\a}}\partial_{\a\da},\quad
\bar{Q}_{\dot{\a}}= -\bar{\partial}^{(\theta)}_{\dot{\a}}
    - i\theta^{\a}\partial_{\a\dot{\a}}, \\
&&S_{\a}=\partial^{(\omega)}_{\a} -
    i\bar{\omega}^{\dot{\a}}\partial_{\a\da} + 2i\theta_{\a}Z,\quad
\bar{S}_{\dot{\a}}=  -\bar{\partial}^{(\omega)}_{\dot{\a}}
    + i\omega^{\a}\partial_{\a\dot{\a}} - 2i\bar{\theta}_{\dot{\a}}\bar{Z}. \nonumber
\end{eqnarray}
Now let us consider the $N=2$, $D=4$ chiral Goldstone superfield
\begin{equation} \plabel{W}
W(x,\theta,\omega) = \phi - \omega^\a W_\a -
    \frac 14\omega^2\bar{D}^2\bar{\phi} -
    i\omega^{\a}{\bar{\omega}}^{\dot{\a}}
    \partial_{\a\dot{\a}} \phi +
    \frac i2 \omega^2 {\bar{\omega}}_{\dot{\a}}\partial^{\a\dot{\a}}W_{\a}-
    \frac 18 \omega^2 {\bar{\omega}}^2\partial^{\a\dot{\a}}
    \partial_{\a\dot{\a}} \phi,
\end{equation}
which obeys Bianchi identities
\begin{equation} \plabel{bi}
D^{(\omega)}{}^2W - \bar{D}{}^2\bar{W} =0,\quad
D^{(\omega)\a}D_{\a}W + {\bar{D}}^{(\omega)}_{\dot{\a}}{\bar{D}}^{\dot{\a}}\bar{W} =0,
\end{equation}
and possesses the following transformation laws
\begin{equation} \plabel{sftrl}
\delta W = c - (\eta^\a S_\a + \bar{\eta}_\da\bar{S}^\da)W,
\quad  Z W = 1
\end{equation}
with respect to the central charge and the
second supersymmetry transformations. Here the standard conventions for the $N=1$ spinor
covariant derivatives are assumed\footnote{The conjugation conditions
are as follows
$$ (\partial_{\a})^\dagger = - \bar{\partial}_{\dot{\a}},\quad
(\partial_{\a\dot{\a}})^\dagger = \partial_{\a\dot{\a}},\quad
(D_{\a})^\dagger =  \bar{D}_{\dot{\a}},\quad
(D_{\a}\bar{D}_{\dot{\a}})^\dagger =- \bar{D}_{\dot{\a}}D_{\a},\quad
(D^{\a}D_{\a})^\dagger = \bar{D}_{\dot{\a}}\bar{D}^{\dot{\a}}. $$}
$$D_{\a} = \partial^{(\theta)}_{\a} - i\bar{\theta}^{\dot{\a}}\partial_{\a\dot{\a}},\quad
\bar{D}_{\dot{\a}} = -\bar{\partial}^{(\theta)}_{\dot{\a}} +
i\theta^{\a}\partial_{\a\dot{\a}}, \quad
D^{(\omega)}_{\a} = D_{\a}\mid_{\theta=\omega}. $$
From the Eqs. \p{W}, \p{sftrl} it follows that the chiral $N=1$
superfields $\phi$, $W_\a$ transform as
\begin{eqnarray} \plabel{vsm}
\delta \phi &=& f + \eta^{\a}W_{\a},\qquad
f = c - 2i\eta^{\a}\theta_{\a}, \\
\delta W_{\a} &=& - \frac12\bar{D}^2\bar{\phi}\eta_{\a} -
2i\partial_{\a\dot{\a}}\phi\bar{\eta}^{\dot{\a}}, \nonumber
\end{eqnarray}
where $W_\a$ is supposed to be constrained by the reality condition
\begin{equation} \plabel{rlc}
D^{\a}W_{\a} + \bar{D}_{\dot{\a}}\bar{W}^{\dot{\a}}=0.
\end{equation}
From Eqs. \p{vsm} we see that owing to the presence of the
inhomogeneous term in the r.h.s. \p{sftrl} the lowest component of
the vector supermultiplet $\phi$ acquires the pure shift term
$f$ in its transformation law. This implies that the superfield
$D_\a\phi$ is shifted on the term proportional to $\eta_\a$ under the
action of the $S$-supersymmetry transformations. Hence, this superfield
can be treated as the Goldstone fermion of spontaneously
broken $N=2$ supersymmetry keeping the $N=1$ supersymmetry unbroken.
This is the typical pattern of \emph{PBGS} theory in which the
corresponding ``longwavelength'' effective action is appeared.
In the most cases these actions can be derived
straightforwardly from covariant constraints imposed on the
supermultiplets incorporating a proper Goldstone multiplet (see, e.g.
\cite{iklz} and refs. therein). In our case, however, this approach
becomes inapplicable because the corresponding linear representation
of the $N=2$, $D=4$ supersymmetry is unknown. Let us try to improve
this situation.

To do this we, first, should solve the constraint \p{rlc} in
terms of $N=1$ gauge real prepotential ${\cal L}$
\begin{equation} \plabel{fs}
W_{\a} = -\frac{i}{4}\bar{D}^2D_{\a}{\cal L},\quad
\bar{W}_{\dot{\a}} = (W_{\a})^\dagger = \frac{i}{4}D^2\bar{D}_{\dot{\a}}{\cal
L},
\end{equation}
and then introduce the ``extended'' vector $N=2$ supermultiplet
\begin{eqnarray} \plabel{L}
\delta {\cal L} &=& f \bar\phi + \bar{f} \phi +
    \eta^\a Z_\a + \bar{\eta}_\da \bar{Z}^\da, \\
\plabel{Z}
\delta Z_\a &=& \bar{f} W_\a - \frac12 \eta_{\a} \bar{D}^2 {\bar F} -
     \bar{\eta}^\da \bar{D}_\da D_\a {\cal L}, \\
\plabel{F}
\delta F &=& f\phi + \eta^\a \Psi_\a, \\
\plabel{Psi}
\delta \Psi_\a &=& f W_\a - \frac12 \eta_{\a} \bar{D}^2{\cal L} -
    2i{\bar{\eta}}^{\da} \partial_{\a\da}F,
\end{eqnarray}
where the auxiliary chiral $N=1$ superfields $F$, $Z_\a$, $\Psi_\a$ are
involved to provide the closure of the transformations \p{L}--\p{Psi}
off-shell. The key point of our investigation is that
the transformations \p{L}--\p{Psi} allow the complex Bianchi identity
\begin{equation}
\plabel{cbi} D^{\a}\Psi_{\a} + \bar{D}_{\dot{\a}}\bar{Z}^{\dot{\a}}=0.
\end{equation}
Keeping into account \p{cbi} one can easily find out that the
transformations \p{L}--\p{Psi}, considered together with \p{vsm}, are closed
off-shell
\begin{equation} \plabel{bo}
[\delta_1,\delta_2]\{{\cal L},~F,~Z_\a,~\Psi_\a\}
= 2i(\eta^\a_1\bar{\eta}^{\dot{\a}}_2
- \eta^\a_2\bar{\eta}^{\dot{\a}}_1)\partial_{\a\dot{\a}}\{{\cal
L},~F,~Z_\a,~\Psi_\a\}.
\end{equation}
The next step is to realize that, apart from the global $N=2$ supersymmetry,
there exist the local $N=2$ supertransformations
\begin{equation} \plabel{gtr}
\delta F = M,\quad \delta \Psi_\a =
\Omega_\a,\quad \delta Z_\a = \Sigma_\a,\quad \delta {\cal L} = N + \bar{N},
\end{equation}
which preserve this supermultiplet. To prove this we should identify
the superfield parameters $\{M$, $N$, $\Omega_\a$, $\Sigma_\a\}$ with
components of a ``complexificated'' $N=2$ vector supermultiplet
\begin{eqnarray} \plabel{cvsm1}
&&\delta M =\eta^{\a}\Omega_{\a},\qquad
\delta \Omega_{\a} = - \frac12\bar{D}^2\bar{N}\eta_{\a} -
2i\partial_{\a\dot{\a}}M\bar{\eta}^{\dot{\a}},\\
&&\delta N = \eta^{\a}\Sigma_{\a},\qquad
\delta \Sigma_{\a} = - \frac12\bar{D}^2\bar{M}\eta_{\a} -
2i\partial_{\a\dot{\a}}N\bar{\eta}^{\dot{\a}},\nonumber
\end{eqnarray}
restricted by the condition
\begin{equation} \plabel{cbi1}
D^{\a}\Omega_{\a} + \bar{D}_{\dot{\a}}\bar{\Sigma}^{\dot{\a}}=0.
\end{equation}
Having at our disposal these parameters one can easily verify that
the redefined superfields
\begin{equation} \plabel{reds}
{\cal L}^\prime = {\cal L}+N+\bar N, \quad
F^\prime = F + M, \quad
\Psi_\a^\prime = \Psi_\a + \Omega_\a, \quad
Z_\a^\prime = Z_\a + \Sigma_\a
\end{equation}
have the same transformation laws as $\{{\cal L}$, $F$, $Z_\a$,
$\Psi_\a\}$.

Now we see, that all these ingredients gathered together indicate
that the set of the
$N=1$ superfields $\{\phi$, ${\cal L}$, $F$, $Z_\a$, $\Psi_\a\}$
actually can be considered as a new off-shell gauge supermultiplet of
the $N=2$, $D=4$ supersymmetry. Note that it appears only in the
framework of the \emph{PBGS} theory. It does not exist beyond this
theory due to the vanishing of the $f$-dependent terms in the r.h.s.
of the Eqs. \p{vsm}, \p{L}--\p{Psi}. It is also very
important to understand that all the subsidiary degrees of freedom $F$,
$\Psi_{\a}$, $Z_{\a}$ as well as those describing the superspin-0
representation of ${\cal L}$ can be finally transformed away owing to
the gauge symmetry \p{gtr}, \p{reds}. Therefore, on the mass-shell we
have the same content of the superfields as that of the vector $N=2$
supermultiplet. Nevertheless, there is an essential difference
between them. In contrast to the vector $N=2$ supermultiplet the new
one includes both the physical superfields $\phi$ and ${\cal L}$
among its components. As we will see in the next section this
significantly simplifies the process of constructing of the corresponding
effective action.

\section{Recursive equation}
In order to do this we follow the standard prescription of refs.
\cite{bg3,rt,iklz} and introduce the recursive equation
\begin{equation} \plabel{req}
{\cal L} = \phi\bar{\phi} + \frac{1}{16}
    \frac{(D^{\a}\phi)(D_{\a}\phi)(\bar{D}_{\da}\bar{\phi})(\bar{D}^{\da}\bar{\phi})}
    {1 -\frac i4 D^{\a}W_{\a} + \frac A4},\qquad
A = - 2(\partial^{\a\da}\phi)(\partial_{\a\da}\phi)
    - \frac 14(D^2\phi)(\bar{D}^2\bar{\phi}).
\end{equation}
Eq. \p{req} completely fixes the gauge freedom of the extended vector
supermultiplet, however, the global $N=2$ supersymmetry remains
unbroken. To prove this we have to come back to the equations
\p{vsm}, \p{L}--\p{Psi}. Analyzing them we observe that
the transformations \p{F}, \p{Psi} admit quite simple nonlinear
solutions
\begin{equation} \plabel{Fc}
F = \frac 12 \phi^2,
\end{equation}
\begin{equation} \plabel{Pc}
\Psi_\a = \phi W_\a.
\end{equation}
Varying these expressions in accordance with \p{vsm} we recover the
transformation laws \p{F}, \p{Psi} and the further constraints
\begin{eqnarray} \plabel{Lc}
{\cal L} &=& \phi\bar{\phi} - \frac i4(W^\a D_\a{\cal L} - Z^\a D_\a \phi), \\
\plabel{Zc}
Z_\a &=& \bar{\phi}W_\a - \frac 12 (\bar{D}^{\dot{\a}}{\cal
L})\partial_{\a\da}\bar{\phi} - \frac
i4(\bar{D}^{\dot{\a}}\bar{\phi})D_\a\bar{D}_{\dot{\a}}{\cal L}.
\end{eqnarray}
Note, that when deriving these ``second--class'' constraints \p{Lc}, \p{Zc}
the Bianchi
identity \p{cbi} must be taken into account. At the first sight these
constraints manifestly contradict the condition of chirality of
the superfield $Z_\a$, which was postulated in the previous section.
In addition, the equation \p{Lc} does not seem to be real as it should
be. However, these discrepancies can be removed simultaneously by the
making use of the ``nilpotence'' conditions following from the
recursive equation \p{req}
\begin{equation} \plabel{fc}
({\cal L} - \phi\bar{\phi})D_\a\phi =
({\cal L} - \phi\bar{\phi})\bar{D}_{\dot{\a}}\bar{\phi} = 0.
\end{equation}
Let us emphasize that these conditions serve as the crucial step
towards the recursive equation \p{req}. Really, as it follows from
\p{fc} the explicit solution of the Eq. \p{Pc}, \p{Zc} can be written
as
\begin{equation} \plabel{ZPc}
Z_\a = - \frac i8 \bar{D}^2D_\a V,\quad
\Psi_\a = - \frac i8 \bar{D}^2D_\a \bar{V},
\end{equation}
where
\begin{equation} \plabel{V}
V = 2 \bar \phi {\cal L} - \phi \bar\phi^2.
\end{equation}
It is easy to see now that for the given $Z_\a$ the recursive
equation \p{Lc} is reduced to its ``canonical'' form \p{req}. So, we
have proved the covariance of the conditions \p{req}, \p{Lc}, \p{Zc}
with respect to $N=2$, $D=4$ supersymmetry. The explicit solution of
the Eq. \p{req} will be obtained in the section \ref{es}. Here,
however, it is instructive to pay our attention to the following
remarkable fact resulting from our considerations.

\section{Infinite-dimensional matrix $N=2$ supermultiplet}
Looking at the Eqs. \p{V}, one cannot pass over a very unexpected
interpretation of this solution. No matter whether there are
composite solutions for
the components of the extended vector supermultiplet or not, in fact,
we always have the possibility to avoid the presence of the fermion
superfields in the supermultiplet we have just constructed. To do
this one should, firstly, incorporate the general $N=1$ superfield $V$
into the content of the supermultiplet. Then, demanding the closure of the
corresponding bracket $N=2$ transformations off-shell we arrive at
the transformation law
\begin{equation} \plabel{nc}
\delta V = 2\bar{f}{\cal L} + 2f\bar{F} -
    \frac i8 \eta^\a \bar D^2 D_\a G +
    \frac i8 \bar\eta_{\dot\a} D^2\bar D^{\dot\a} H,
\end{equation}
where
\begin{equation} \plabel{GH}
G=2\phi^2 {\cal L} - \frac43 \phi^3 \bar\phi, \qquad
H=4\phi\bar\phi{\cal L} - 3\phi^2\bar\phi^2.
\end{equation}
Repeating this process step by step we end up with the following
infinite-dimensional matrix $N=1$ superfield
\begin{equation} \plabel{Vmn}
V_{mn} = V^\dag_{nm} = \frac{2^{m+n-2}}{m!n!}\left(mn\phi^{m-1} \bar\phi^{n-1} {\cal L}
    +(1-mn)\phi^m \bar\phi^n\right),\qquad  m,n \geq 0.
\end{equation}
One can check that this superfield transforms as follows
\begin{eqnarray}\plabel{infdim}
\delta V_{mn} &=& 2f V_{m-1,n} + 2 \bar f V_{m,n-1}-
    \frac i8 \eta^\a \bar D^2 D_\a V_{m,n+1} +
    \frac i8 \bar\eta_{\dot\a} D^2\bar D^{\dot\a} V_{m+1,n},\\
\delta V_{0n} &=& 2\bar f V_{0n-1}+
    \frac i8 \bar\eta_{\dot\a} D^2\bar D^{\dot\a} V_{1n}, \nonumber
\end{eqnarray}
when $\phi$ and ${\cal L}$ are varied in accordance with Eqs. \p{vsm},
\p{L}, \p{ZPc}, \p{V}. Moreover,
examining the transformation laws \p{infdim} we reveal that they are closed
off-shell (just in accordance with the superalgebra \p{alg}) for all the
components $V_{mn}$ treating as new \emph{independent} $N=1$ superfields
when the corresponding matrix elements are normalized as follows
\begin{equation} \plabel{Vm}
\pmatrix{V_{00} & V_{01} & V_{02} &   & \ldots\cr
         V_{10} & V_{11} & V_{12} & V_{13} &  \cr
         V_{20} & V_{21} & V_{22} &   &  \cr
           & V_{31} &   &   &  \cr
         \vdots &   &   &  & \ddots   } =
\pmatrix{1/4& \bar\phi/2 & \bar F && \ldots\cr
         \phi/2 & {\cal L} & \bar V & \bar G &  \cr
         F & V & H & &  \cr
           & G &   & &  \cr
         \vdots &   &   &   & \ddots},
\end{equation}
with $V_{n0}$ being chiral.
Thus, beginning with either the finite-dimensional $N=2$ vector
supermultiplet \p{vsm} or its extended analog \p{L}--\p{Psi}
collecting both the \emph{bosonic} and \emph{fermionic} components
we necessarily
finish with the infinite-dimensional $N=2$ matrix supermultiplet
$V_{mn}$ composed of the \emph{bosonic} components only.
It is obvious, that by the definition
$V_{mn}$ proves to be infinite-reducible and can be
``truncated'' by the spinor
covariant derivatives up to the vector $N=2$ supermultiplet \p{vsm}.
Its advantage, however, is that it immediately leads to the extended
vector $N=2$ supermultiplet which gives the algorithmic
procedure of constructing the ``canonical'' form of the recursive
equation \p{req}. Otherwise, it may be merely guessed. Hence,
regarding the super-3-brane itself, we have proved the existence of
some new fundamental representation of the \emph{PBGS} theory which
the effective Lagrangian density along with the corresponding
chiral Goldstone superfield belongs to.

Notice, that earlier a very resembling procedure was demonstrated in
the framework of $N=4$, $D=4$ \emph{PBGS} theory \cite{bik}. There
the $N=2$, $D=4$ Born--Infeld superfield Lagrangian density was
revealed among the components of an infinite-dimensional off-shell
supermultiplet providing \emph{PBGS}. Just as $V_{mn}$ this
supermultiplet encompasses the infinite set of the worldvolume
superfields obeying an infinite set of covariant constraints. In
contrast to our case these constraints are solved there
only iteratively.

\section{Lagrangian density} \label{es}
A closed form of the Lagrangian density for the supersymmetric
3-brane can be established uniquely from the recursive equation
\p{req} rewritten preliminary in the following more convenient form
\begin{equation} \plabel{req1}
L = \frac{1}{16}\frac{(D\phi)^2(\bar{D}\bar{\phi})^2}
    {1 - \frac{1}{16} D^{\a}\bar{D}^2D_{\a}L + \frac A2}, \quad
L \equiv {\cal L} - \phi \bar{\phi}.
\end{equation}
This equation can be solved exactly since the numerator of \p{req1}
contains the maximal number of the Grassmann spinors $D_\a \phi$,
$\bar{D}_{\da}\bar{\phi}$. Therefore in the denominator we should
leave only the terms which do not include these spinors. Thus we get
the following ``effective'' equation
\begin{equation} \plabel{efeq}
X \equiv  D^\a \bar D^2 D_\a L = \frac{1}{16}
\frac{\Big(D^\b\bar{D}^2D_\b
(D\phi)^2(\bar{D}\bar{\phi})^2\Big)_{eff}}{1 + \frac A2 - \frac{X}{16}},
\end{equation}
where
\begin{eqnarray} \plabel{num}
\Big(D^\b\bar{D}^2D_\b(D\phi)^2(\bar{D}\bar{\phi})^2\Big)_{eff} &=&
(D^2\phi)^2(\bar{D}^2\bar{\phi})^2 +
64(\partial^{\a\da}\phi)(\partial_{\a\da}\phi)
(\partial^{\s\ds}\bar{\phi})(\partial_{\s\ds}\bar{\phi})+\nonumber\\
&+&16(D^2\phi)(\bar{D}^2\bar{\phi})
(\partial^{\a\da}\phi)(\partial_{\a\da}\bar{\phi})
\end{eqnarray}
Solving \p{efeq} and choosing the solution with the nonsingular
behaviour in the limit $\phi = 0$
\begin{equation} \plabel{X}
X = 8 \left(1 + \frac A2 - \sqrt{1 + A + B}\right), \qquad
\end{equation}
$$
B = (\partial^{\a\da}\phi)(\partial_{\a\da}\bar{\phi})
(\partial^{\s\ds}\phi)(\partial_{\s\ds}\bar{\phi})
- (\partial^{\a\da}\phi)(\partial_{\a\da}\phi)
(\partial^{\s\ds}\bar{\phi})(\partial_{\s\ds}\bar{\phi}),
$$
we finally find the following expression for the Lagrangian density
\begin{equation} \plabel{Ld}
{\cal L} = \phi \bar{\phi} + \frac 18
\frac{(D^{\a}\phi)(D_{\a}\phi)(\bar{D}_{\da}\bar{\phi})
(\bar{D}^{\da}\bar{\phi})}{1 + \frac A2 + \sqrt{1 + A + B}}.
\end{equation}
This is the canonical form of the Lagrangian density for the super-3-brane
obtained in a framework of another approach in refs. \cite{bg2,rt}.
Note that the corresponding action
\begin{equation} \plabel{act}
S = \int d^4xd^4\theta {\cal L},
\end{equation}
is manifestly invariant with respect to the unbroken supersymmetry
transformations while under the broken one it is shifted in accordance
with Wess--Zumino theory on the surface term
$\int d^4xd^4\theta (f \bar\phi + \bar{f} \phi +
\eta^\a Z_\a + \bar{\eta}_\da \bar{Z}^\da).$

\section{Conclusions}
Thus, we have shown that the most natural way of description of the
supersymmetric 3-brane is achieved when both the chiral Goldstone $N=1$
superfield and the Lagrangian density are embedded
into the infinite-dimensional matrix $N=2$ supermultiplet $V_{mn}$
restricted by the conditions \p{req}, \p{Vmn}. It is proved that he
latter appears in the framework of \emph{PBGS} theory and solves
ultimately the problem of constructing of the supersymmetric 3-brane
action in its closed form.

Nevertheless, many problems remain outside this work. One of them is
the problem of connection of this approach with the nonlinear
realization of the $N=2$, $D=4$ supersymmetry in the superspace
\cite{bg1,ks}. Having ensured such a connection we will have an
opportunity to test the general principle of the Nambu--Goldstone
theories: the uniqueness of the Goldstone interaction arising in
models with various mechanisms of spontaneous breaking of
supersymmetry. As to the system considered here it would be desirable
to understand how the chiral Goldstone multiplet of the linear
realization $\phi$ is interlinked with that of the nonlinear
realization \cite{bg1}, and how a universal closed form action of
this realization looks like?

Related question is whether this approach will allow us to solve a
problem of spontaneous breaking of $D=6$ Lorentz symmetry in target
space. From the general reasons we know that the part of this
symmetry related to the transverse directions should be realized
nonlinearly, in Goldstone mode fashion \cite{bg1}. Unfortunately, up
to now we do not know how the corresponding Goldstone worldvolume
superfields are involved into the residual \emph{PBGS} action.

At last, there is the most intrigue question: whether the \emph{PBGS}
theory we have considered here can be promoted to the case of $AdS_6$
background, as it has been done, for example, for the
$N=1$, $D=4$ supermembrane in ref. \cite{dik}?

We hope to answer these questions in future.

We are grateful to E. Ivanov, S. Krivonos, A. Pashnev and B. Zupnik for
kind hospitality at
Bogoliubov Laboratory of Theoretical Physics, JINR, where this work
has begun.

\def\NPB #1 #2 #3{{\sl Nucl. Phys.} {\bf B #1} (#2) #3}
\def\PRD #1 #2 #3{{\sl Phys. Rev.} {\bf D #1} (#2) #3}
\def\PRev #1 #2 #3{{\sl Phys. Rev.} {\bf  #1} (#2) #3}
\def\PLB #1 #2 #3{{\sl Phys. Lett.} {\bf B #1} (#2) #3}
\def\JPA #1 #2 #3{{\sl J. Physics} {\bf A#1} (#2) #3}
\def\SJNP #1 #2 #3{{\sl Sov. J. Nucl. Phys. (Yadern.Fiz.)} {\bf #1} (#2) #3}

\end{document}